\documentstyle[11pt,epsfig,menu97]{article}

\newcommand{\beq}{\begin{equation}}
\newcommand{\eeq}{\end{equation}}
\newcommand{\beqa}{\begin{eqnarray}}
\newcommand{\eeqa}{\end{eqnarray}}

\begin{document}
    \setlength{\baselineskip}{2.6ex}

\title{EFFECTIVE FIELD THEORY FOR THE TWO--NUCLEON SYSTEM}
\author{Ulf-G. Mei\ss ner\\
{\em FZ J\"ulich, IKP (Th), D-52425 J\"ulich, Germany}}

\maketitle

\vspace{-2.5cm}

\hfill {\small FZJ-IKP(TH)-1999-23}

\vspace{1.9cm}

\begin{abstract}
\setlength{\baselineskip}{2.6ex}
\noindent I discuss the dynamics of the two--nucleon system as obtained
from a chiral nucleon--nucleon potential. This potential is based on a modified
Weinberg power counting and contains chiral one-- and two--pion exchange
as well as four--nucleon contact interactions. The description of the
S--waves is very precise. Higher partial waves are also well reproduced.
We also find a good description of most of the deuteron properties.

\end{abstract}

\setlength{\baselineskip}{2.6ex}

\section*{INTRODUCTION}

\noindent One of the most important and most intensively studied
problems of nuclear physics is deriving the forces between nucleons.
While there have been many successful and very precise models (of more
or less phenomenological
type), only in the last decade the powerful methods of effective
field theory (EFT) have been used to study this question. In
particular, Weinberg~\cite{weinNN} employed power--counting to the
irreducible n--nucleon interaction and obtained leading order results
by iterating such type of potential in a Lippmann--Schwinger equation.
This type of resummation is necessary to deal with the shallow bound
states (or large S--wave scattering lengths) present in the
two--nucleon system. This is in contrast to conventional chiral
perturbation theory in the meson and meson--nucleon sectors, where
all interactions can be treated perturbatively.
A full numerical study based on Weinberg's approach at
next-to-next-to-leading (NNLO) order was performed in ref.\cite{ubi}.
It was concluded that the approach could give qualitative insight but
was not precise enough to compete with the accurate modern potentials
or even phase shift analysis. In addition, a novel power counting scheme 
was proposed by Kaplan, Savage and Wise (KSW)~\cite{ksw}. In that
approach, only the leading order momentum--independent four--nucleon
interaction is iterated and all other effects, in particular the
coupling of pions, are treated perturbatively. This is in stark
contrast to Weinberg's scheme, were one--pion exchange (OPE) is
present already at leading order and is iterated (among other interactions).
What I will show in the following is that a suitably modified Weinberg
scheme can be turned into a precision tool, which allows to study
systematically the interactions between few nucleons. This lends credit
to Weinberg's ideas spelled out almost 10 years ago.

\section*{CHIRAL EXPANSION OF THE NN POTENTIAL}

\noindent EFT is, by construction, only useful in a space of momenta
below a certain scale. The latter depends on the system one is
investigating. In what follows, I will be concerned with the
effective potential between nucleons as defined (and somewhat
modified) in ref.\cite{weinNN}.
In the EFT approach suggested by Weinberg, one has to
deal with two different types of interactions. First, there is OPE, 
two--pion exchange and so on, to describe the long and medium range
physics. Second, there are four--nucleon contact interactions to
describe the short (and to some extent the medium) range physics.
So we have a scale separation, the dividing line being somewhere
above twice the pion mass and below the typical scale of chiral
symmetry breaking, $\Lambda_\chi \simeq 1\,$GeV. The problem at hand
can be treated exactly by integrating out the pionic degrees of
freedom from the Fock space using the projection
formalism of Okubo, Fukuda, Sawada and Taketani~\cite{OFST}. The
usefulness of this approach when applied to momentum space has been
demonstrated in a toy--model calculation, see refs.\cite{egkm}.
Based on this approach, we have set up the following scheme. First,
one constructs the irreducible chiral NN potential based on a power
counting in harmony with the projection formalism. This is outlined
in detail in ref.\cite{egm}. To third order in small momenta, this
potential is given by the following contributions (LO = leading order,
NLO = next-to-leading order):
\begin{description}
\item[LO] OPE with lowest order insertions and two 4N contact interactions
          without derivatives.
\item[NLO] Vertex and self--energy corrections to the LO interactions,
           TPE with lowest order insertions and seven 4N contact
           interactions with two  derivatives.
\item[NNLO] Vertex and self--energy corrections to OPE as well as TPE with exactly
  one insertion from the dimension two $\pi N$ Lagrangian. These terms encode non--trivial
  information about the pion--nucleon interaction beyond leading order and are thus
  sensitive to the chiral structure of QCD.
\end{description}
This potential is divergent. All divergences (of quadratic and
logarithmic form)  can
be dealt with by subtracting divergent loop integrals, which leads
to an overall renormalization of the axial--vector coupling $g_A$ and
seven of the nine coupling constants related to the 4N interactions. The
precise prodecure is discussed in detail in ref.\cite{egmNN}. The
renormalized potential still
has a bad high energy behaviour. Some of the
contact interactions (NNLO TPE contributions) grow quadratically 
(cubically) with increasing momenta.  Even the momentum--independent
contact interactions necessitate regularization, which is performed
on the level of the Lippmann--Schwinger equation. That is done in the following way:
\beq\label{reg}
V( \vec{p},\vec{p}~'\,) \to f_R ( \vec{p}\,) \, V( \vec{p},\vec{p}~'\,) \,
f_R (\vec{p}~'\,)~,
\eeq
where $f_R ( \vec{p}\,)$ is a regulator function chosen in harmony with the
underlying symmetries. In ref.\cite{egmNN}, two different regulator
functions are used, the sharp cutoff  $f_R^{\rm sharp} ( \vec{p}\,) = 
\theta (\Lambda^2 -p^2)~,$ and an exponential form, $f_R^{\rm expon} (
\vec{p}\,) = \exp(-p^{2n} / \Lambda^{2n})~,$
with $n = 2,3,\ldots\,$. The latter form is more suitable for the
calculation of some observables. Bound and scattering states can then
be obtained by solving the Lippmann--Schwinger equation with the regularized
potential.

\section*{PARAMETERS AND FITTING PROCEDURE}

\noindent In this section, I briefly describe how the parameters
are pinned down. The parameters related to the pion--nucleon interaction
beyond leading order can be fixed by a fit of the chiral perturbation
theory pion--nucleon amplitude~\cite{fms} to the dispersion--theoretical one
inside the Mandelstam triangle~\cite{paul}.
In addition, we have nine coupling constants related to four--nucleon
contact interactions. These can be uniquely determined by a fit to the S-- and
P--waves together with the mixing parameter $\epsilon_1$. While both S--waves
contain two parameters,  the P--waves and $\epsilon_1$ depend one (more precisely,
one can form linear combinations of  the LECs which appear as these parameters
in the considered partial waves). At NLO, the resulting values for the
LECs related to the 4N contact terms are sensitive to the energy range used in the fit.
At NNLO, the resulting values are more stable, so that we
can perform global fits for energies up 100~MeV. For example, such a global fit
with $\Lambda = 500$~MeV at NLO and 875~MeV at NNLO leads to a deuteron binding
energy of $E_d = -2.17$ and $-2.21$~MeV at NLO and NNLO, respectively. Therefore,
without any fine tuning, we can reproduce the empirical value within 2 percent
and 1 permille at NLO and NNLO, in order. The increase of the cut--off
value when going from NLO to NNLO is related to the fact that at NNLO, the
chiral TPEP includes mass scales above the two--pion mass, which is the scale
related to the uncorrelated TPEP appearing at NLO. It is also worth
mentioning that the
quality of the fits increases visibly as one goes from LO to NLO to NLO (for details,
see ref.\cite{egmNN}). This is, of course, expected from the
underlying power counting.
With that, one can predict these partial waves for energies above 100~MeV. All other
partial waves with angular momentum $\ge 2$ and the deuteron properties are {\it
predictions}.

\subsection*{Prediction for the S-- and P--waves}
 
\noindent The resulting S--waves are shown in figs.1,2 in comparison to the
Nijmegen phase shift analysis (PSA)~\cite{nijpwa}. The improvement when going
from LO to NLO to NNLO is clealy visible.

\parbox{7cm}{
\begin{center}
\epsfig{figure=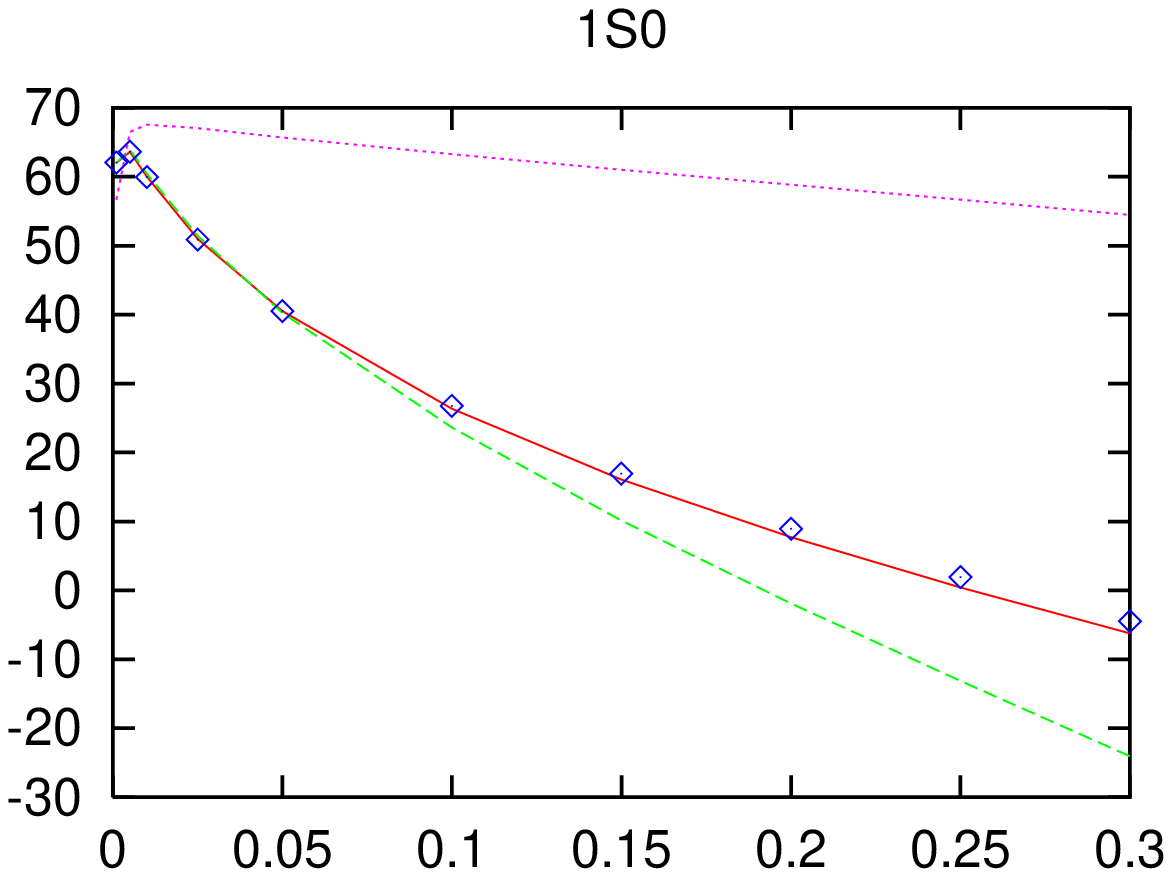,width=6.5cm,height=5cm}
\end{center}}
\parbox{6.2cm}{\vspace*{1.7cm}
{\small \setlength{\baselineskip}{2.6ex} Fig.~1. Predictions for
  the $^1S_0$ partial wave (in degrees) at LO (purple curve), NLO 
  (green curve)  and NNLO (red curve) in
  comparison to the Nijmegen PSA (blue diamonds) for nucleon laboratory
  energies up to 0.3~GeV.
}}

\parbox{7cm}{
\begin{center}
\epsfig{figure=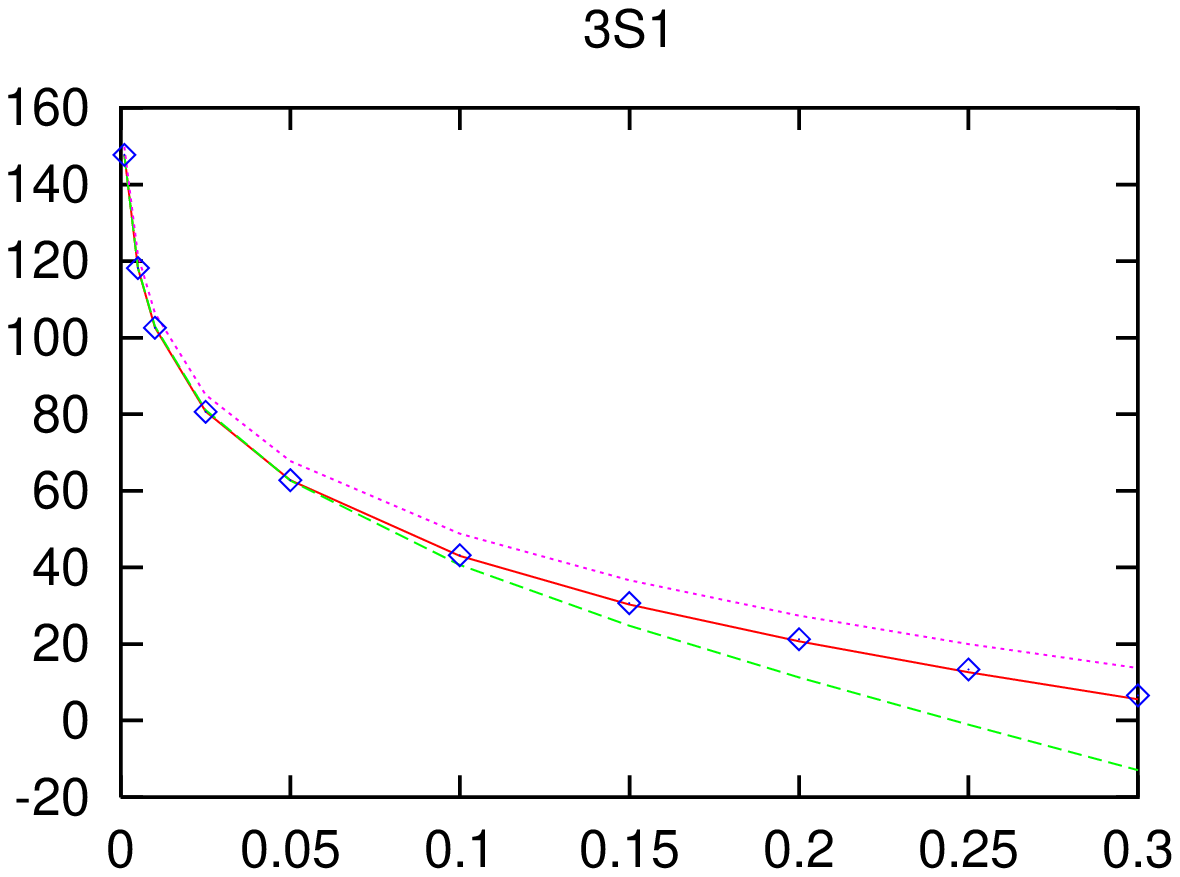,width=6.5cm,height=5cm}
\end{center}}
\parbox{6.2cm}{\vspace*{1.7cm}
{\small \setlength{\baselineskip}{2.6ex} Fig.~2. Predictions for
  the $^3S_1$ partial wave (in degrees) at LO (purple curve), g
  NLO (green curve)  and NNLO (red curve) in
  comparison to the Nijmegen PSA (blue diamonds) for nucleon laboratory
  energies up to 0.3~GeV.
}}

\noindent The P--waves are mostly well described, although the NNLO TPEP is 
a bit too strong
in $^3P_1$ and $^3P_2$. Of particular interest is $\epsilon_1$ since it has also
been calculated at NLO~\cite{ksw} and NNLO~\cite{calt} in the KSW approach.
Our results (note that we used the so--called Stapp parametrization~\cite{stapp}
for the coupled triplet waves)
 are shown in comparison to the ones of refs.\cite{ksw,calt} in
fig.3 as a function of nucleon cms momentum up to 350~MeV. For energies below
150~MeV, the KSW results are comparable to ours, but is obvious from that figure
that their approach is tailored to work at low energies. If one wants to go to momenta
above 100~MeV, it appears that pion exchange should be treated non--perturbatively.

\parbox{8cm}{
\begin{center}
\epsfig{figure=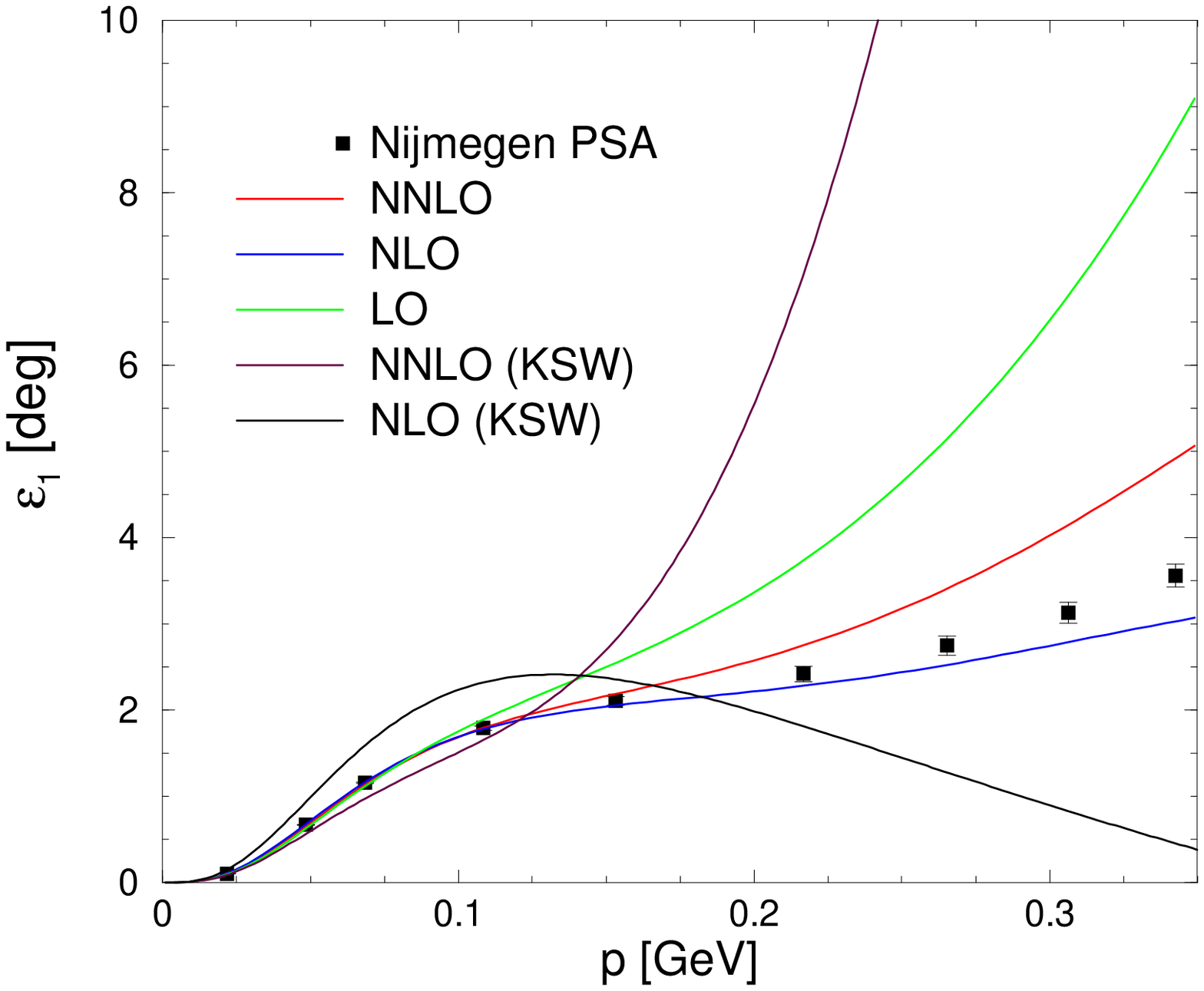,width=7cm,height=6cm}
\end{center}}
\parbox{5.2cm}{\vspace*{3cm}
{\small \setlength{\baselineskip}{2.6ex} Fig.~3. The $^3S_1-^3D_1$ mixing
  parameter $\epsilon_1$ for our approach and the KSW scheme in comparison
  to the Nijmegen PSA as a function of the nucleon cms momentum.}}

\noindent I also would like to discuss briefly the so--called effective
range expansion. For any partial wave, one can write
\beq\label{ERE}
p\, \cot\delta = -\frac{1}{a} + \frac{1}{2}\, r \, p^2
+ v_2 \, p^4 + v_3 \,p^6 + {\cal O}(p^{8})~,
\eeq
with $\delta$ the phase shift, $p$ the nucleon cms momentum, 
$a$ the scattering length and
$r$ the effective range. It has been stressed in ref.\cite{CH}
that the shape parameters $v_i$ are a good testing ground for the range
of applicability of the underlying EFT. At NNLO, we find e.g.
$a = 5.424 \,(5.420)\,$fm, $r = 1.741~(1.753)\,$fm, $v_2 = 0.046~(0.040)\,$fm$^3$,
and $v_3 = 0.67~(0.67)\,$fm$^5$ for $^3S_1$. Similarly, for $^1S_0$, we have
$a = -23.72 \,(-23.74)\,$fm, $r = 2.68~(2.67)\,$fm, $v_2 = -0.61~(-0.48)\,$fm$^3$,
and $v_3 = 5.1~(4.0)\,$fm$^5$. The numbers in the square
brackets refer to the $np$ system from the Nijmegen~II potential . 
Note that one can also perform the fit
such that the scattering lenghts and effective ranges are exactly
reproduced. This leads only to modest changes in the values of
$v_{2,3}$, e.g. for such a fit in $^1S_0$ one has $v_2 
= -0.53\,$fm$^3$ and $v_3 = 5.0\,$fm$^5$.
This rather good agreement illustrates again that the long range
physics associated with pion exchanges is incorporated
correctly and it demonstrates the predictive power of such an EFT approach.

\subsection*{Predictions for higher partial waves}

\noindent Consider first the D-- and F--waves. These are free of parameters
and most problematic since the NNLO TPEP can be too strong. In some potential
models, TPEP is simply cut at distances of (approximately) less than one fermi. 
Nevertheless,
we find a rather satisfactory description of these partial waves. Of particular
interest is $^3D_1$ since it is related to the deuteron channel. Also, $^1D_2$
is supposedly very sensitive to contributions from the $\Delta$--resonance,
which in our approach is subsumed in the LECs related to the dimension two
$\pi N$ interaction. Both these partial waves are well described, see figs.4,5.
Even the small $^3D_3$ partial wave is very well described up to the opening
of the pion production threshold (in the Bonn potential, this partial
wave is dominated by correlated TPE). Note, however, that the D--waves
are very sensitive to the choice of the regulator cut--off.

\parbox{7cm}{
\begin{center}
\epsfig{figure=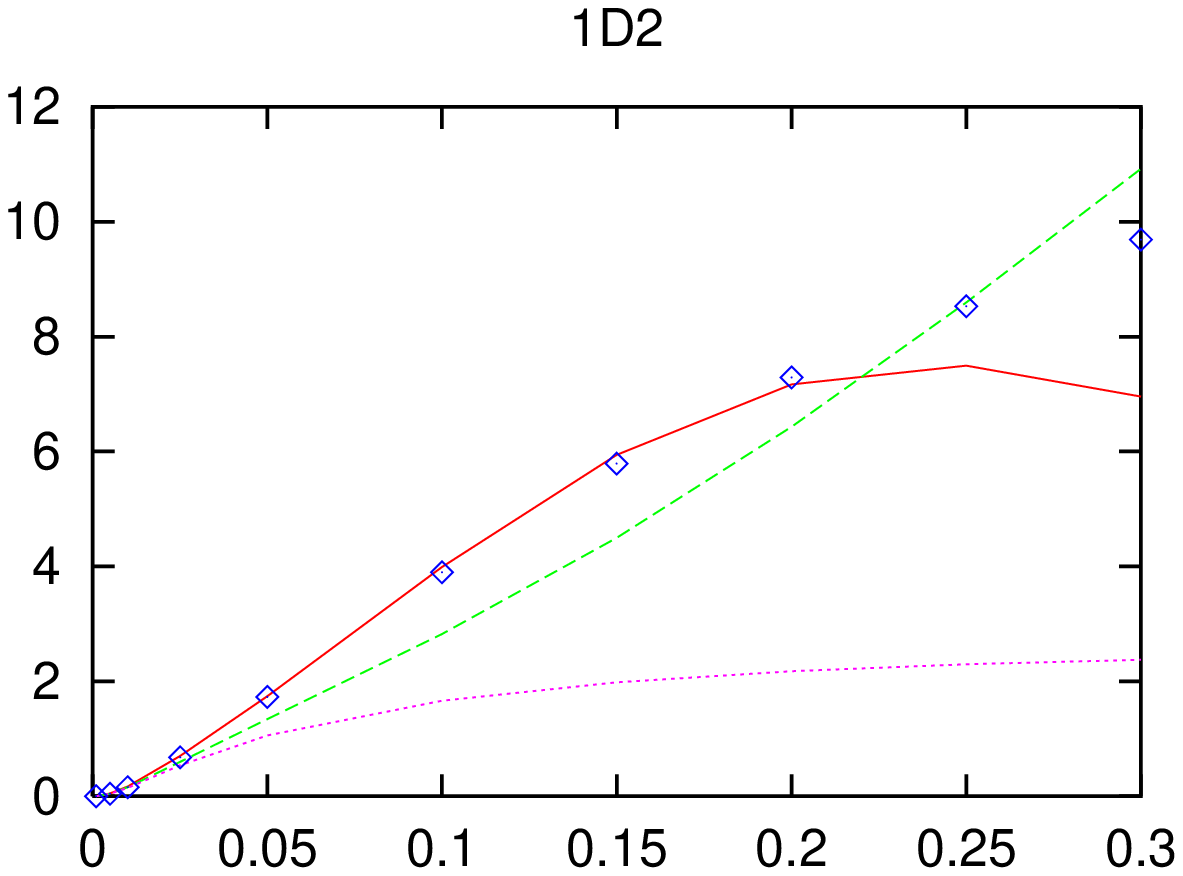,width=6.5cm,height=5cm}
\end{center}}
\parbox{6.2cm}{\vspace*{1.7cm}
{\small \setlength{\baselineskip}{2.6ex} Fig.~4. Predictions for
  the $^1D_2$ partial wave (in degrees) at LO (purple curve), NLO 
  (green curve)  and NNLO (red curve) in
  comparison to the Nijmegen PSA (blue diamonds) for nucleon laboratory
  energies up to 0.3~GeV.
}}

\noindent The so--called peripheral waves ($l \ge 4$) have already
been considered by the Munich group~\cite{norb}. Their calculation
was based on Feynman graphs using dimensional regularization. The 
potential was constructed perturbatively by proper partial wave projection
of the NNLO OPE and TPE. While
in most of the peripheral waves OPE is dominant, there are a few
exceptions where chiral NNLO TPEP is needed to bring the predictions in
agreement with the data or PSA results. Our calculation, which is based
on a completely different regularization scheme and treats the potential
non--perturbatively, leads to the same results. This is rather gratyfying.
One particular example that demonstrates the importance of NNLO TPEP 
is $^3G_5$ as shown in  fig.6.

\parbox{7cm}{
\begin{center}
\epsfig{figure=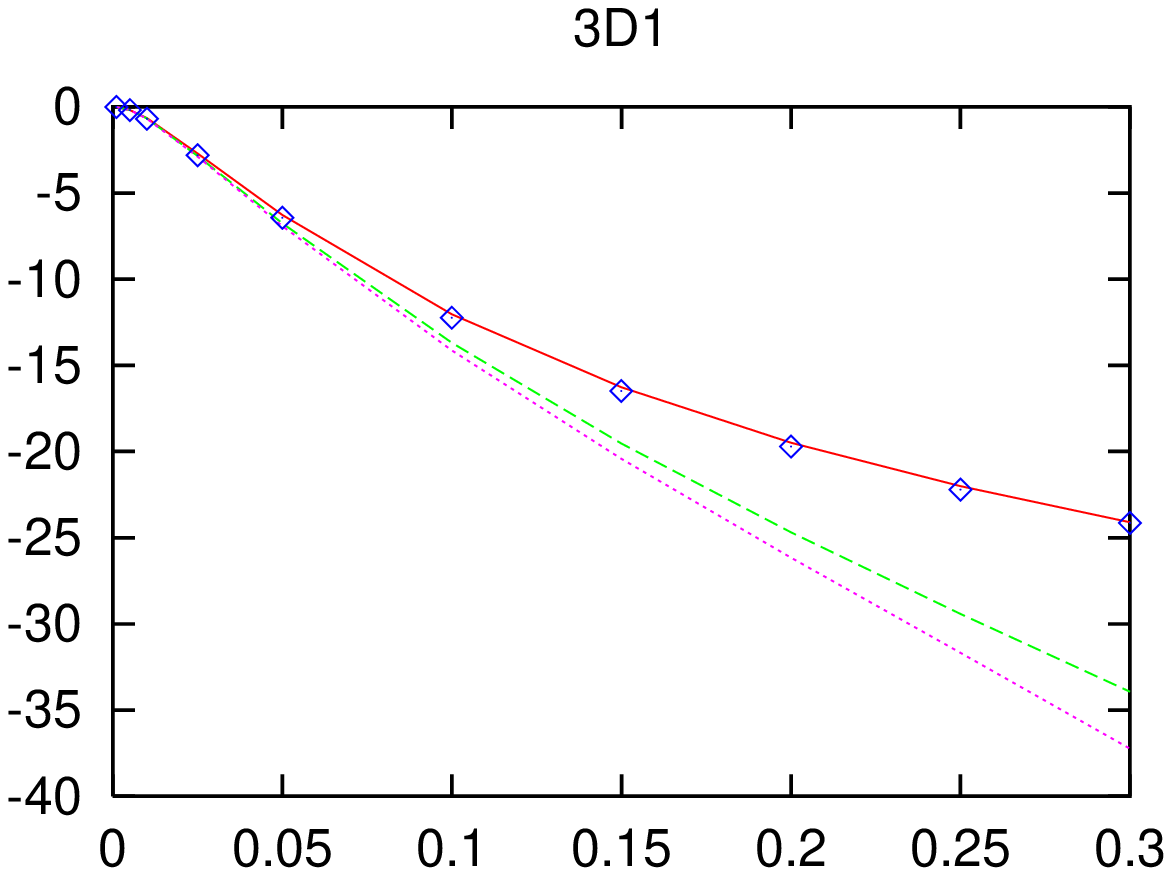,width=6.5cm,height=5cm}
\end{center}}
\parbox{6.2cm}{\vspace*{1.7cm}
{\small \setlength{\baselineskip}{2.6ex} Fig.~5. Predictions for
  the $^3D_1$ partial wave (in degrees) at LO (purple curve), g
  NLO (green curve)  and NNLO (red curve) in
  comparison to the Nijmegen PSA (blue diamonds) for nucleon laboratory
  energies up to 0.3~GeV.
}}

\parbox{7cm}{
\begin{center}
\epsfig{figure=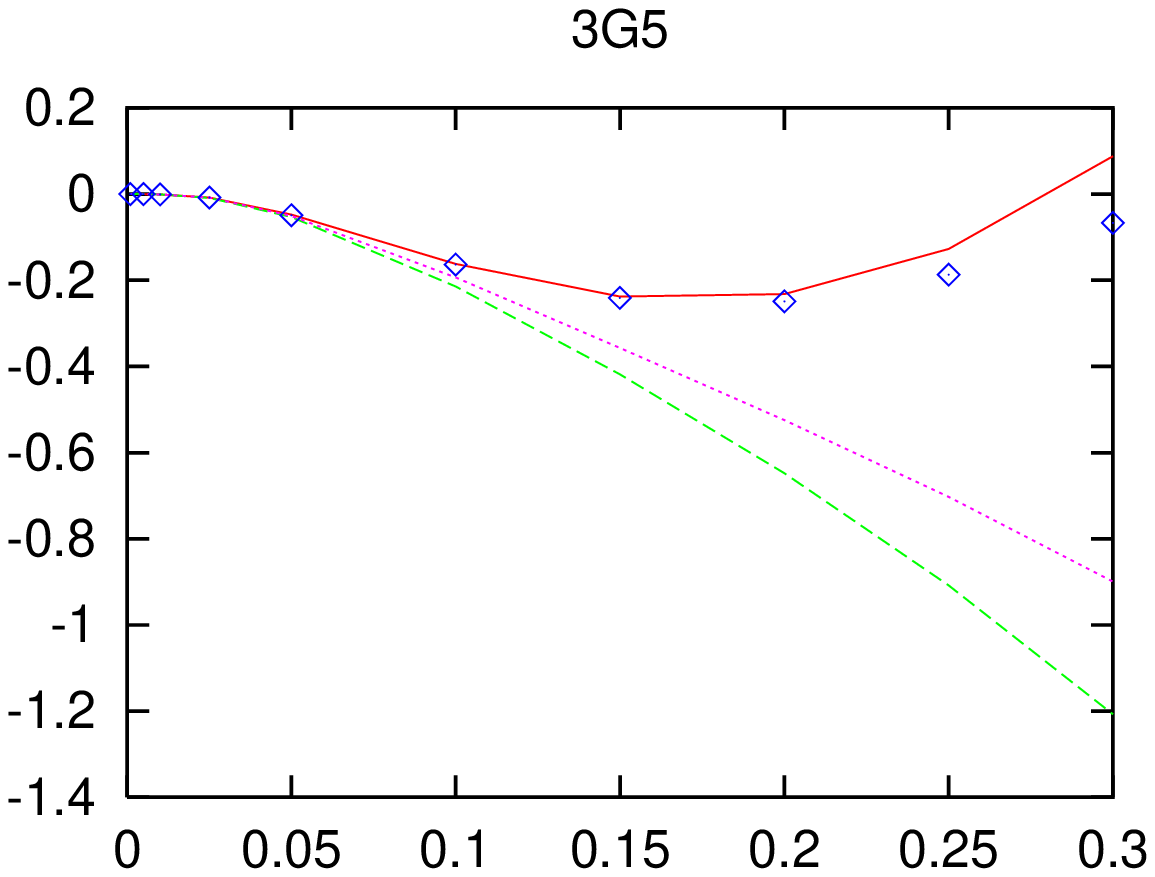,width=6.5cm,height=5cm}
\end{center}}
\parbox{6.2cm}{\vspace*{1.7cm}
{\small \setlength{\baselineskip}{2.6ex} Fig.~6. Predictions for
  the $^3G_5$ partial wave (in degrees) at LO (purple curve), 
  NLO (green curve)  and NNLO (red curve) in
  comparison to the Nijmegen PSA (blue diamonds) for nucleon laboratory
  energies up to 0.3~GeV.
}}

\subsection*{Deuteron properties}
\noindent
It is straightforward to calculate the bound state properties. 
At NNLO (NLO), we use an exponential
regulator with $\Lambda =1.05\,(0.60)\,$GeV, which reproduces the deuteron
binding energy within an accuracy of about one third of a permille
(2.7 percent). No attempt is made to reproduce this number with better precision.
In table~1 the deuteron properties are collected and compared to the
data and two realistic potential model predictions.
\begin{table}[bht] 
\begin{center}
\caption{Deuteron properties derived from our chiral potential
    compared to two ``realistic'' potentials
    (Nijmegen--93~\cite{Nij93}
    and  CD--Bonn~\cite{cdbonn}) 
    and the data. Here, $r_d$ is the
    root--mean--square matter radius. An exponential regulator
    with $\Lambda = 600\,$MeV and $\Lambda = 1.05\,$GeV 
    at NLO and NNLO, in order, is used. \label{tab:D}}
\vspace{0.2cm}
\begin{tabular}{||l||c|c||c|c||c||}
    \hline
    & NLO  &  NNLO  &  Nijm93 & CD-Bonn & Exp.    \\ 
    \hline   \hline  
$E_d$ [MeV] & 2.1650  &2.2239 & 2.224575 & 2.224575 & 2.224575(9) \\
    \hline
$Q_d$ [fm$^2$] & 0.266  & 0.261 & 0.271 & 0.270 & 0.2859(3) \\
    \hline
$\eta$ & 0.025   & 0.025  & 0.0252 & 0.0255 & 0.0256(4)\\
    \hline
$r_d$ [fm] & 1.975  & 1.967 & 1.968 & 1.966 & 1.9671(6)\\
    \hline
$A_S$ [fm$^{-1/2}$] & 0.866  & 0.887 & 0.8845 & 0.8845 & 0.8846(16)\\
    \hline
$P_D [\%]$ & 3.8   & 6.5  & 5.67 & 4.83 &  -- \\
    \hline
  \end{tabular}
\end{center}

\end{table}
\noindent
We note that deviation of our prediction for the quadrupole
moment compared to the empirical value slightly larger than for the
realistic potentials.  Still, it remains to be checked whether this
problem persists when one includes the meson--exchange currents
(compare also the discussion in ref.\cite{V18}). The asymptotic $D/S$ ratio,
called $\eta$, and the strength of the asymptotic wave function, $A_S$, are well
described. The D--state probability, which is not an observable, is
most sensitive to small variations in the cut--off. At NLO, it is comparable
and at NNLO somewhat larger than obtained in the CD-Bonn or the
Nijmegen-93 potential. This increased value of $P_D$ is related to the
strong NNLO TPEP. At N$^3$LO, I expect this to be compensated by
dimension four counterterms. It is also worth mentioning that at NNLO,
we have two additional very deeply bound states. These have, however,
no influence on the low--energy physics and can be projected
out. Furthermore, these states are an artefact of the too strong potential
and will most probably vanish at N$^3$LO. Altogether, the description of the
deuteron as compared to ref.\cite{ubi} is clearly improved. 

\subsection*{COORDINATE SPACE REPRESENTATION}
\noindent
It is also illustrative to consider the coordinate space
representation of this potential. I point out that it is
intrinsically non--local in  momentum as well as in coordinate
space. Therefore, it cannot be directly compared to standard local
NN potentials. In fig.7, the corresponding potential in the $^1S_0$
partial wave at NLO is shown. Qualitatively, it exhibits all expected features,
namely the short--range repulsion, intermediate range attraction and 
dominance of pion exchange at large separations (as much as this
can be seen in a pictorial of a non--local potential). In ref.\cite{egmNN},
a more detailed comparison of the chiral potential with the so--called
realistic potentials is given.

\parbox{9cm}{
\vspace{-1cm}
\begin{center}
\epsfig{figure=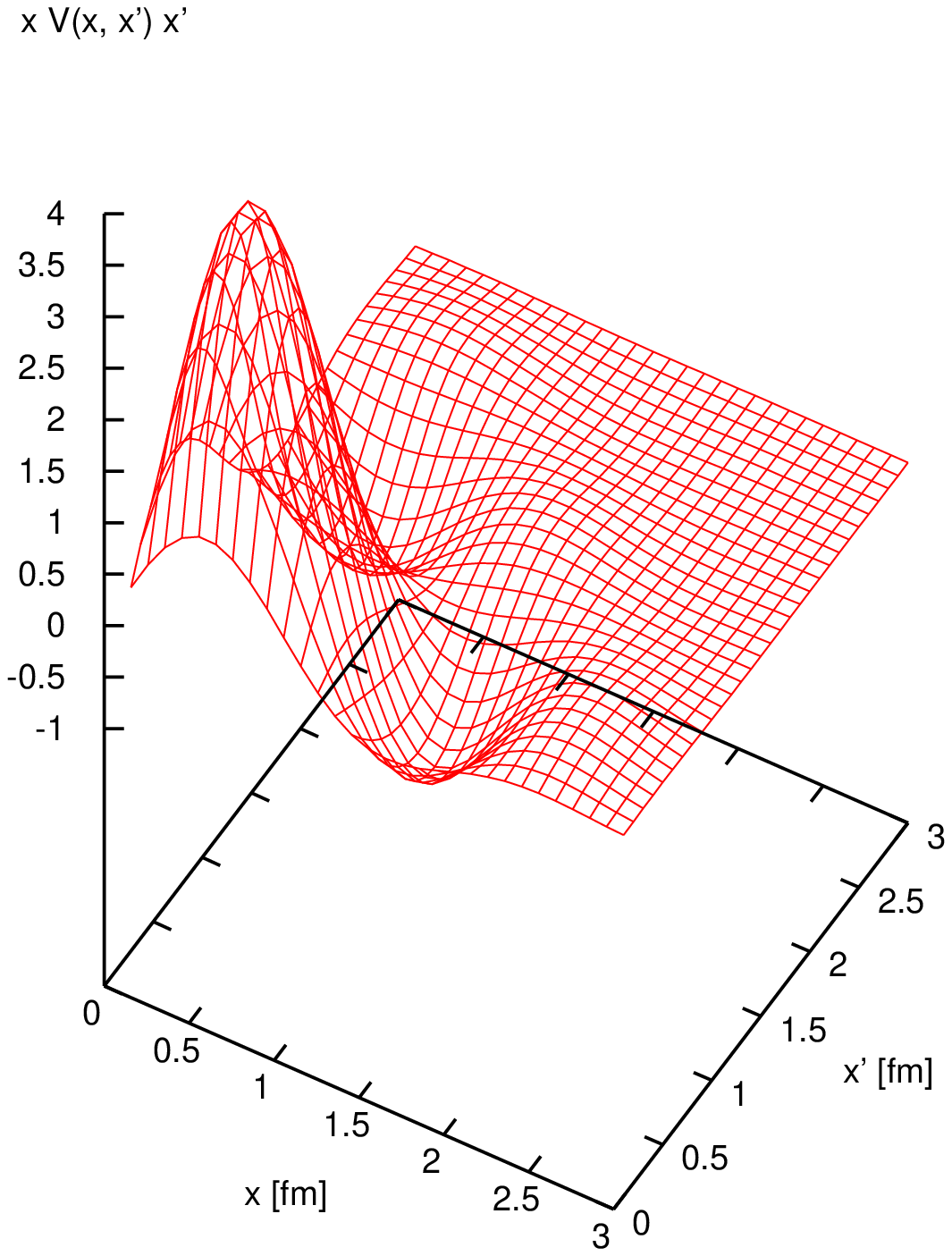,width=8.5cm,height=8.cm}
\end{center}}
\parbox{4.2cm}{\vspace*{4.3cm}
{\small \setlength{\baselineskip}{2.6ex} Fig.~7. Coordinate space
  representation of the potential $x V(x,x')x'$ in the 
  the $^1S_0$ partial wave at NLO.
}}

\section*{SUMMARY AND OUTLOOK}

\noindent In this talk, I have shown some results for the
nucleon--nucleon interaction based on effective field theory.
The formalism is an extension of the ideas spelled out by 
Weinberg almost a decade ago. The power counting is performed on the
level of the potential. We have worked out the
potential to next-to-next-to-leading order in the power counting.
At NNLO, it consists of 
one-- and two--pion exchange diagrams with insertions from
the dimension one and two pion--nucleon Lagrangian. The corresponding
low--energy constants from ${\cal L}_{\pi N}^{(2)}$
have been determined from a fit ot the $\pi N$ amplitudes inside
the Mandelstam triangle~\cite{paul}. In addition, there are two/seven 
four--nucleon contact interactions with zero/two derivatives.
The so--defined potential is divergent. All these divergences can
be absorbed by a proper redefinition of the axial--vector coupling
$g_A$ and of seven of the nine four--nucleon couplings.
This renormalized potential needs to be regularized due to its bad
high momentum behaviour, see eq.(\ref{reg}). The regularized potential
is used
in a Lippmann--Schwinger equation to obtain bound and scattering
states by menas of standard Gauss--Legendre quadrature.
The corresponding coupling constants can be obtained after proper
partial wave decomposition from a fit of the two S-- and four P--waves
as well as the $^3S_1 - ^3D_1$ mixing parameter $\epsilon_1$ 
for nucleon lab energies below 100~MeV. The resulting S--waves are
as accurate as obtained from high precision potentials (for energies
up to the pion production threshold). The
other partial waves are mostly well described. For angular momentum
$\ge 2$, all phase shifts are parameter free predictions. In
particular, the $^3D_1$ wave is well reproduced. In some of the D--
and F--waves, the NNLO TPEP is somewhat too strong. This will be cured
at N$^3$LO due to the appearance of 4N contact interactions with four
derivatives. Their contribution is expected to balance the short
distance contribution from the TPEP. In most peripheral waves ($l \ge 4$), 
OPE is dominant but there are some exceptions where NNLO TPE
is needed to close the gap between the EFT prediction and the data
(or PSA). This has been already found before in refs.~\cite{norb,bras} 
based on  completely different approaches. In addition, without any
fine tuning we obtain good results for the deuteron, the sole
exception being the too low quadrupole moment. However, it is
mandatory to calculate the pertinent exchange currents before drawing
a conclusion on this issue. In fact, the EFT approach can easily be
extended to the coupling of external fields as well as to systems
with more than two nucleons. It is my opinion that the chiral Lagrangian approach
does more than ``$\ldots$ \, justify approximations (such as assuming
the dominance of two--body interactions) that have been used for many
years by nuclear physicists\ldots''~\cite{weinD}.

\section*{ACKNOWLEDGEMENTS}

\noindent I am grateful to Evgeny Epelbaum and Walter  Gl\"ockle for
a most enjoyable collaboration and allowing me to present these results
before publication. I would like to thank Iain Stewart for supplying
me with the results for $\epsilon_1$ in the KSW scheme and Vincent
Stoks for providing the range parameters. Last but not
least the superbe organization by Christine Kunz and Res Badertscher
is warmly acknowledged.

\bibliographystyle{unsrt}

\end{document}